# Electric Charge Quantization in Standard Model

O. B. Abdinov, F.T. Khalil-zade, S. S. Rzaeva

Institute of Physics of Azerbaijan National Academy of Sciences
AZ143, Baku, G. Javid av.33

**Abstract.** In the framework of Standard Model for the arbitrary values of Higgs and fermions fields hypercharges, taking into account parity invariance of electromagnetic interaction, expressions for the fermions charges, testifying the electric charge quantization are obtained. From the chiral anomalies cancellation condition within one family of leptons and quarks, numerical values of fermions charges, coinciding with standard values of charges have been obtained.

The Standard Model (SM) of the strong and electroweak interactions [1] is an excellent description of the interactions of elementary particles down to distance $\sim 10^{-16}$ sm. Though the SM is a good phenomenological theory and coincides very well with all experimental result [2], there are theoretical difficulties which suggest that SM should be an effective model at low energies. Some of these difficulties are: the existence of three families, the mass hierarchy problem, CP violation, the quantization of an electric charge etc. Attempts for solution of these difficulties were made in [3-5]. The SU(5) grand unification model unifies the interactions and predicts the electric charge quantization. The mass hierarchy problem has found its solution in supersymmetric and superstring theories. The model based on $E_6$ group symmetry unifies the electroweak and strong interactions and may explain the masses of the neutrinos. It should be notice, that some of these difficulties were reached, or by enlarging the group of symmetry, or introducing a large particle content.

However, as it will be shown further, in the present work, one of theoretical difficulties of SM - electric charge quantization problem can be solved and within the framework of SM.

Let's consider Weinberg – Salam (WS) model for one family of leptons and quarks. To complete the analysis we will assume that neutrino has got the right component. In this case in the model we have the following fermionic fields

$$\psi_L = \begin{pmatrix} v_L \\ e^- \end{pmatrix}_L, \quad \psi_{eR} = e_R, \quad \psi_{vR} = v_R, \quad \psi_{QL} = \begin{pmatrix} u \\ d \end{pmatrix}_L, \quad \psi_{uR} = u_R, \quad \psi_{dR} = d_R, \quad (1)$$

and Hiqqs isodoublet

$$\varphi = \begin{pmatrix} \varphi^+ \\ \varphi^0 \end{pmatrix}. \quad (2)$$

(Neutrinos and quarks mixing are not considered.)

Interaction lagrangian of fermions and Hiqqs field with gauge bosons in WS model looks li

$$L = i\bar{\psi}_L D_\mu \psi_L + i\bar{\psi}_{eR} D_\mu \psi_{eR} + i\bar{\psi}_{vR} D_\mu \psi_{vR} + i\bar{\psi}_{QL} D_\mu \psi_{QL} + \\ i\bar{\psi}_{uR} D_\mu \psi_{uR} + i\bar{\psi}_{dR} D_\mu \psi_{dR} + (D_\mu \varphi)^* (D_\mu \varphi), \quad (3)$$

where

$$D_\mu = \partial_\mu - ig\vec{T}\vec{A} - ig'\frac{Y}{2} B_\mu. \quad (4)$$

For hypercharges $Y$ of Hiqqs (2) and fermionic fields (1) we choose following designations

$$Y(\varphi) = y, \quad Y(\psi_L) = y_L, \quad Y(\psi_{eR}) = y_{eR}, \quad Y(\psi_{vR}) = y_{vR}, \\ Y(\psi_{QL}) = y_{QL}, \quad Y(\psi_{uR}) = y_{uR}, \quad Y(\psi_{dR}) = y_{dR}. \quad (5)$$

Suppose that hypercharges (5) are real. As usually, for isoscalar fields with $T = 0$, the second term in (4) is zero, and for fields $\varphi, \psi_L, \psi_{QL}$ - $\vec{T} = \vec{\tau}/2$.

Transformation of fields $A_\mu^3$ and $B_\mu$ to the physical ones $A_\mu$ и $Z_\mu$ looks like

$$A_\mu^3 = A_\mu \sin\theta_N + Z_\mu \cos\theta_N,$$
$$B_\mu = A_\mu \cos\theta_N - Z_\mu \sin\theta_N. \tag{6}$$

Taking into account (5) we have

$$\sin\theta_N = yg'/\bar{g}_N, \cos\theta_N = g/\bar{g}_N, \bar{g}_N = \sqrt{g^2 + y^2 g'^2}. \tag{7}$$

Let's consider interaction of leptons with the electromagnetic field. From (3), (5) and (6), we obtain

$$L_{l\gamma} = \bar{\nu}\gamma_\mu (Q_\nu + Q'_\nu \gamma_5)\nu A_\mu + \bar{e}\gamma_\mu (Q_{0e} + Q'_{0e}\gamma_5)e A_\mu. \tag{8}$$

where

$$Q_\nu = \frac{g}{4}\left(1 + \frac{y_L + y_{\nu R}}{y}\right)\sin\theta_N, \quad Q'_\nu = \frac{g}{4}\left(1 + \frac{y_L - y_{\nu R}}{y}\right)\sin\theta_N,$$
$$Q_{0e} = -\frac{g}{4}\left(1 - \frac{y_L + y_{eR}}{y}\right)\sin\theta_N, \quad Q'_{0e} = -\frac{g}{4}\left(1 - \frac{y_L - y_{eR}}{y}\right)\sin\theta_N. \tag{9}$$

From the expression (8) one can see that the interaction of neutrino with a photon differs from zero and there are terms proportional $\gamma_5$ in neutrino – photon and electron – photon interactions. Taking into account the parity invariance of electromagnetic interaction, from (9) we obtain

$$y_L = y_{\nu R} - y, \quad y_L = y_{eR} + y. \tag{10}$$

Substituting values (10) in expressions (9) leads

$$Q_\nu = \frac{Q_e}{2}\left(1 + \frac{y_L}{y}\right), \quad Q_{0e} = -\frac{Q_e}{2}\left(1 - \frac{y_L}{y}\right), \tag{11}$$

where $Q_e = g\sin\theta_N$.

If $Q_\nu = 0$ from (11) we have

$$y_L = -y, \quad Q_{0e} = -Q_e. \tag{12}$$

Notice, that in the case when neutrino has not the right component ($\psi_{\nu R} = 0, y_{\nu R} = 0$, SM with massless neutrino), the requirement parity invariance of electromagnetic interaction and the condition of neutrino charge equality to zero are equivalent. In the same time expressions (10) testifies the hypercharge conservation in leptons mass lagrangian $f_e \bar{\psi}_L \psi_{eR}\varphi = m_e \bar{e}e$, $f_\nu \bar{\psi}_L \psi_{\nu R}\varphi^c = m_\nu \bar{\nu}\nu$, accordingly.

For the interaction of quarks with electromagnetic field we have

$$L_{q\gamma} = \bar{u}\gamma_\mu (Q_{1u} + Q'_{1u}\gamma_5)u A_\mu + \bar{d}\gamma_\mu (Q_{2d} + Q'_{2d}\gamma_5)d A_\mu. \tag{13}$$

where

$$Q_{1u} = \frac{g}{4}\left(1 + \frac{y_{QL} + y_{uR}}{y}\right)\sin\theta_N, \quad Q'_{1u} = \frac{g}{4}\left(1 + \frac{y_{QL} - y_{uR}}{y}\right)\sin\theta_N,$$
$$Q_{2d} = -\frac{g}{4}\left(1 - \frac{y_{QL} + y_{dR}}{y}\right)\sin\theta_N, \quad Q'_{2d} = -\frac{g}{4}\left(1 - \frac{y_{QL} - y_{dR}}{y}\right)\sin\theta_N. \tag{14}$$

Similarly, to interaction of leptons with an electromagnetic field, from the condition of parity invariance of electromagnetic interaction from (14), we obtain for quarks

$$Q'_{1u} = 0, \quad Q'_{2d} = 0, \tag{15}$$

Relations (14) and (15) lead to the conditions

$$y_{QL} = y_{uR} - y, \quad y_{QL} = y_{dR} - y. \tag{16}$$

Conditions (16) correspond to hypercharge conservation in quarks mass lagrangian
$f_u \bar{\psi}_{QL} \psi_{uR} \varphi^c = m_u \bar{u}u$ and $f_d \bar{\psi}_{QL} \psi_{dR} \varphi = m_d \bar{d}d$, accordingly.

From (15) and (14) we have

$$Q_u = \frac{Q_e}{2}\left(1 + \frac{y_{QL}}{y}\right), \quad Q_d = -\frac{Q_e}{2}\left(1 - \frac{y_{QL}}{y}\right). \tag{17}$$

The obtained expressions (11) and (17) can be considered as the evidence of electric charge quantization of leptons and quarks. However the expressions (17) do not define numerical values of quarks charges (in terms of electron charge).

For this purpose let's consider cancellation of chiral anomalies [6]. In the considered case for the cancellation of $\gamma_5$ - anomalies within the framework of one family of leptons and quarks should be satisfied the following relation

$$y_L = -3y_{QL}. \tag{18}$$

Taken into account the expressions (18) and (10) in (17) we obtain

$$Q_u = \frac{2}{3}Q_e, \quad Q_d = -\frac{1}{3}Q_e. \tag{19}$$

Note that the similar to (11) and (17) expressions can be obtained and for other families of leptons and quarks. These results define quantization and numerical values of leptons and quarks electric charge.

In a considered case, expressions for charges (11) and (17) can be written in a following general form

$$\frac{Q_f}{Q_e} = T_{3L}^f + \frac{Y_f}{2}. \tag{20}$$

where $T_{3L}^f$ - the third isotopic spin component, and, $Y_f = Y_{fL}/y$, $Y_{fL}$ - hypercharge of left izomultiplets. It should be noticed that the expression (20), coinciding formally with known Gell-Mann-Nishijima formula is the generalization of the obtained expressions for the electric charge of particles. This expression as well as Gell-Mann-Nishijima formulae describe the charges of particles, but unlike the last explain the values of particles charges (taking into account the conditions between izomultiplets hypercharges).

In the Standard Model and various extended models of electroweak interaction photon eigenstate does not contain vacuum averages of Higgs fields, but depends on the hypercharges of Higgs fields. It leads in turn to dependence of electron charge from hypercharges of Higgs fields. Dependence of a charge of particles from hypercharges of Higgs fields leads to the conclusion that the Higgs fields influence to the "formation" of a particles charges. Thus, Higgs fields are responsible not only for occurrence of particles mass, but also for the formation of their charges. Certainly, the further and detailed studies of this problem are necessary.

## Conclusion

1. There is an electric charge quantization in the Standard Model, though in the hidden form.
2. The formula (19) for fermions charges, obtained in this work and similar formally to the Gell-Mann-Nishijima one, describes particles charge quantization.
3. Formulas (11), (16) for fermions charges with taking into account the chiral anomalies cancellation condition, explain fractional quarks charge values.
4. Dependence of the particles charge from the Higgs field's hypercharges, can be interpreted as the new property of Higgs fields. Higgs fields are responsible not only for occurrence particles mass, but also for the formation of their charges.